\documentclass[twocolumn,showpacs,showkeys]{revtex4}
\usepackage{graphicx,subfigure,epsfig}
\def\beq{\begin{equation}}
\def\eeq{\end{equation}}
\def\bea{\begin{eqnarray}}
\def\eea{\end{eqnarray}}
\def\d{{\mathrm{d}}}

\newfont{\cursive}{pzcmi at 9pt}
\def\~t{\tilde{t}}
\def\Painleve{Painlev\'{e}}

\def\e2phi{\e^{2\Phi}}

\begin{document}

\title{Black Holes without Event Horizons}

\author{Alex B. Nielsen} \email{eujin@phya.snu.ac.kr}
\affiliation{Center for Theoretical Physics
and School of Physics\\ College of Natural Sciences
Seoul National University, Seoul 151-742, Korea  \\}

\begin{abstract}
We discuss some of the drawbacks of using event horizons to define black holes. The reasons are both practical, physical and theoretical. We argue that locally defined trapping horizons can remedy many of these drawbacks. We examine of the question of whether black hole thermodynamics should be associated with event horizons or trapping horizons. To this end we discuss what role trapping horizons may play in black hole thermodynamics. In addition, we show how trapping horizons may give rise to Hawking radiation and discuss the issue of gravitational entropy.

\end{abstract}
\pacs{04.70.-s, 04.70.Bw, 04.70.Dy}
\keywords{black holes, Hawking radiation, trapping horizons, gravitational entropy}
\maketitle

\section{Introduction}

Black holes have come to play an important role in physics. In astrophysics, they represent the end point of stellar collapse for sufficiently large stars. A great number of likely stellar-sized black hole candidates have already been observed. Supermassive black holes seem to occur in most galaxies and appear to play an important role in active galactic nuclei and quasars. It is possible that supermassive black holes have an important role to play in galaxy formation. Black hole mergers also represent one of the most promising candidates for observable gravitational wave sources with the new generation of gravitational wave detectors.

From a theoretical viewpoint, the importance of black holes is perhaps even greater. Ever since the original results on black hole uniqueness and black hole thermodynamics, black holes have been used as testing grounds for ideas about quantum gravity and possible hints as to the form such a theory should take. It is often claimed that one of the greatest triumphs of string theory is its ability to reproduce the Bekenstein-Hawking area-entropy relation from the counting of string microstates. A great deal is now known about black holes in higher dimensions, black holes in lower dimensions, black holes in higher derivative gravity theories, black holes coupled to various matter fields and black holes in non-trivial backgrounds.

Clearly there are a great number of interesting physical phenomena in which black holes are expected to play some role. But what exactly is a black hole? There are two separate possibilities for defining a black hole. Either one could try to define a black hole in terms of some geometrical property of spacetime or one could define a black hole in terms of the global causal structure of spacetime. In General Relativity both geometry and causal structure are important, although they are logically distinct.

For many years black holes have been defined theoretically in terms of event horizons. That is, in terms of the past causal boundary of some region of spacetime, usually future null infinity. The black hole region is defined as that part of spacetime that is not in the causal past of future null infinity and is therefore bounded by the event horizon. This is very much a definition based on causal structure. This definition is well established and supported by a range of arguments \cite{astresocclus}.

However, it is possible that the definition of a black hole in terms of an event horizon is not the most useful definition for many of the physical phenomena listed above. Here we will argue that this is indeed the case. There are a variety of reasons for this, both practical, physical and theoretical. We will suggest that black holes may be far more usefully defined in terms of trapping horizons and will show what relation trapping horizons may have to the question of black hole thermodynamics.

The difference between black holes defined in terms of event horizons and black holes defined in terms of trapping horizons will be most acute in the case of dynamically evolving black holes. Many models of astrophysical phenomena assume some background black hole spacetime such as Kerr or Schwarzschild and consider perturbative processes on this background. The distinction will not make much difference in these cases, since for the Kerr and Schwarzschild spacetimes the event horizon is also a trapping horizon. However, in truly dynamical situations such as black hole formation or black hole merger simulations there is likely to be some difference. Perhaps most importantly, in the case of evaporating black holes, the difference may be crucial.

In the first section we will discuss some of the drawbacks of event horizons. We will then, in section two, illustrate how trapping horizons can be easily located in a spherically symmetric spacetimes. In the third section we will show how the familiar laws of black hole dynamics can be derived using trapping horizons and in section four we will indicate what role trapping horizons may play in Hawking radiation. We will conclude with some remarks about gravitational entropy and some speculation on implications for black hole thermodynamics.

\section{Event horizons}

Event horizons are usually defined as the past causal boundary of future null infinity. This definition captures the idea of causal signals being unable to `escape'. It also naturally entails that causal signals cannot be sent from inside the event horizon to any point outside the horizon. In general, this definition will depend on the choice of region for which one wants to calculate the causal past. If one wants that region to be `at infinity' then event horizons depend critically on the spacetime structure all the way to infinity.

The telelogical nature of the definition means that in some sense event horizons `know' about the future. Their dynamical evolution reacts to processes that may not even have registered in the past light cone yet. As such, the definition is highly non-local. If there were a large enough distant shell collapsing down on us, there could be an event horizon passing through us right now. Because of this large collapsing shell it may be that light signals we send out now cannot reach infinity.

A related feature of event horizons is that they can, in principle, arise and evolve in exactly flat regions of spacetime. Consider a hollow spherically symmetric thin shell of matter, with mass $M$, collapsing under its own gravity in an otherwise vacuum spacetime. By Birkhoff's theorem we know that the exterior of the shell is a portion of Schwarzschild space and the interior of the hollow shell is exactly flat Minkowski space. An observer sitting at the centre of the shell can imagine firing radially-outgoing photons. These photons will move outwards across Minkowski space and after some time will meet the collapsing shell. Before the collapsing shell of matter has passed within its own Schwarzschild radius ($r=2M$) the photons will be able to pass through the shell and escape to infinity. If the photon reaches the shell just as the shell passes through $r=2M$ then the photon will be trapped, along with all subsequent photons. This photon's trajectory will form part of the event horizon.

Therefore, the event horizon will come into existence in purely flat space and it's area will increase at the speed of light until it reaches the surface $r=2M$. Notice also that the increasing area of the event horizon is not caused by any matter flowing over it instantaneously, but rather by the future `anticipation' of infalling matter.

It is well known that event horizons are difficult to locate in numerical simulations. Locating event horizons in dynamical simulations is notoriously difficult, that is to say time consuming (see for example \cite{Thornburg:2006zb}). Perhaps the easiest way is to propagate null lines back from infinity and hope that they asymptote to the event horizon. For this to work, finding event horizons in numerical solutions also requires a solution that is stable all the way to `infinity', or at least until it settles down to an approximately stationary state. It is far easier numerically to locate apparent horizons on a given hypersurface and in many cases, use this as a `proxy' for the event horizon.

Event horizons do serve several useful purposes in numerical codes. In excision methods, the interior of the event horizon represents the maximal region that can be excised without influencing the future development of the exterior region. It is in this sense that using the apparent horizon as a proxy is most useful since, for most dynamical simulations of say black hole merger, any apparent horizon will also lie inside the event horizon. As an aside, here we note that the apparent horizon will lie inside the event horizon as long as the null energy condition is satisfied. This is a reasonable assumption for astrophysical modeling but will break down when quantum effects are taken into account through Hawking radiation.

Another use of event horizons is in comparing different numerical codes. Since the location of the event horizon is absolute, independent of the space-time slicing used to generate the solution, its location, if it can be reliably found, can be used as a diagnostic to compare different simulations using different foliations. In these respects, event horizons serve as useful practical tools if they can be found. 

However, there are other drawbacks of event horizons of a more physical nature. An obvious drawback is that it is impossible to locate an event horizon using local measurements. That is to say, it is impossible to locate an event horizon with the tools available to finite, mortal physicists. One needs to know the entire future of the universe. This means that it is impossible to test experimentally whether an event horizon even exists and therefore impossible to test whether black holes truly exist. The existence of event horizon defined black holes is technically beyond the scope of experimental verification! Even if one passed over an event horizon, classically one would not notice.

One could argue that for all practical purposes, such and such an object was practically spherically symmetric with a practically vacuum exterior and therefore described by the Schwarzschild metric. One could then measure the mass and areal radius of such an object by the deviation of test masses and conclude that there was, for all practical purposes, an event horizon at $r=2M$. However, these approximations would only ever be approximately true, especially if the object was embedded in some expanding universe with cosmic microwave background and gravitational waves. The object would also only be static as long as one ignored the far distant future when it might evaporate. It is the telelogical nature of the definition of black holes that causes this problem. Whether one would be able to perform a quantum mechanical experiment that would reveal the existence of an event horizon is a question we would like to address.

It would seem that black hole thermodynamics does not even require an event horizon. Various authors have been successful in deriving dynamical laws for locally defined dynamical and trapping horizons \cite{Hayward:1993wb,Ashtekar:2004cn}. These laws are analogous to the usual laws of black hole thermodynamics. In this context, it is important to remember that event horizons do not necessarily coincide with trapping horizons. While many event horizons can be given the structure of event horizons, there are certainly many situations where trapping horizons are not event horizons. If thermodynamical relations can be derived for two different types of horizon then the question arises, which one, if any, represents the `true' thermodynamic system?

It also seems likely that event horizons are not required for Hawking radiation. This is perhaps not surprising since one would like to believe that a local quantum field theory on a curved spacetime would only depend on locally defined structures. Since both quantum field theory and general relativity are local field theories, it would be very surprising if non-local behaviour could arise from their combination. That is not to say that a putative theory of quantum gravity cannot give rise to non-local effects. Merely that, in semi-classical gravity, which is presumably all one needs to study the Hawking radiation process, one would not expect non-local structures to play a role.

To see some of these theoretical considerations more clearly and to give a useful example of how trapping horizons can be used, we turn now to an example of trapping horizons in spherically symmetric spacetimes.

\section{Trapping horizons in spherical symmetry}

As an alternative to event horizons, one may consider defining the black hole as the region inside a trapping horizon. The idea of a trapping horizon is based on the notion of a marginal surface. In a four dimensional spacetime with Lorentzian signature, every two dimensional spacelike surface has two null normals associated with it, that are unique up to rescalings. A marginal surface is a two dimensional spacelike surface for which the expansion, $\theta$, of one of the null normals to the surface vanishes.

The expansion can be thought of as measuring whether neighbouring light rays are being focused or defocused by the gravitational field. A positive $\theta$ refers to defocusing, a negative $\theta$ to focusing. In fact, the expansion represents the behaviour of a circle drawn on the spacelike two surface as it is instantaneously propagated along one of the null directions. If the light rays are being focused the area of this circle will be decreasing and $\theta$ will be negative.

The two null normals will be denoted here by $n^{a}$ and $l^{a}$ and will we refer to them as the ingoing and outgoing null directions respectively. Basically they represent the instantaneous path followed by light rays attempting to escape from the surface and the vanishing of the expansion of one of the null normals means that light traveling in this direction is instantaneously neither focused nor defocused by the geometry.

It is important to realise that this requirement does not mean that light rays cannot move away from the surface and indeed, as soon as they leave the surface, they are, in principle, free to move outwards and `expand'. It is only instantaneously at the surface that the expansion is required to be zero.

More formally, for a spacelike two-surface with null normals $n^{a}$ and $l^{a}$ (such that $n^{a}l_{a} = -1$), the expansion associated with the vector $l^{a}$ can be computed by
\beq \label{expansion} \theta_{l} = g^{ab}\nabla_{a}l_{b} + n^{a}l^{b}\nabla_{a}l_{b} + l^{a}n^{b}\nabla_{a}l_{b} \eeq
with a similar form for $\theta_{n}$. A trapping horizon, more properly a future outer trapping horizon, is defined by Hayward \cite{Hayward:1993wb} as the closure of a three-surface which is foliated by marginal surfaces, for which $\theta_{l}=0$, and which, in addition, satisfies
\begin{enumerate}
\item[\it{i}.] $\theta_{n}<0$ (to distinguish between white holes
and black holes).
\item[\it{ii}.] $n^{a}\nabla_{a}\theta_{l}<0$ (to distinguish
between inner and outer horizons of, for example, the non-extremal Reissner-Nordstr\"{o}m solution).
\end{enumerate}\bigskip
To show how these definitions can be applied in a simple situation we turn now to an example. Any spherically symmetric metric in four dimensions can be put in the form \cite{Nielsen:2005af}
\bea \d s^2 & = & - e^{-2\tilde{\Phi}(t,r)}\left(1-\frac{2m(t,r)}{r}\right)\d t^{2} + \nonumber \\ & & \frac{\d r^{2}}{\left(1-\frac{2m(t,r)}{r}\right)}+r^{2}\d\Omega^{2} \eea
in so-called Schwarzschild or curvature coordinates, where $m(t,r)$ is immediately recognisable as the Misner-Sharp mass function. As is well known, these coordinates are undefined at the points $r=2m(r,t)$. A better coordinate system for examining the behaviour in this region are the \Painleve -Gullstrand coordinates
\bea \d s^{2} & = & -e^{-2\Phi(\tau,r)}\left(1-\frac{2m(\tau,r)}{r}\right)\d \tau^{2}+ \nonumber \\ & & 2e^{-\Phi(\tau,r)}\sqrt{\frac{2m(\tau,r)}{r}}\d\tau\d r + \d r^{2} + r^{2}\d\Omega^{2} \eea
The radial null geodesics for this metric can be easily found by setting $\d s = \d\Omega = 0$. For this we find
\beq \label{dtdtau} \frac{\d r}{\d\tau} = -e^{-\Phi(\tau,r)}\left( 1\pm \sqrt{\frac{2m(\tau,r)}{r}}\right) \eeq
where the plus sign denotes the ingoing geodesics. Thus we can find outgoing geodesics $l^{a}$ and ingoing geodesics $n^{a}$ with components
\beq \label{l} l^{a} = \left(e^{\Phi(\tau,r)},1-\sqrt{\frac{2m(\tau,r)}{r}},0,0\right) \eeq
\beq \label{n} n^{a} = \frac{1}{2}\left(e^{\Phi(\tau,r)},-1-\sqrt{\frac{2m(\tau,r)}{r}},0,0\right) \eeq
The factor of two ensures that the cross normalisation is the conventional $n^{a}l_{a} = -1$. Then, using (\ref{expansion}) we can compute
\beq \theta_{l} = \frac{2}{r}\left(1-\sqrt{\frac{2m(\tau,r)}{r}}\right) \eeq
\beq \theta_{n} = -\frac{1}{r}\left(1+\sqrt{\frac{2m(\tau,r)}{r}}\right) \eeq
We see that the expansion of $n^{a}$ is always negative and that at $r=2m(\tau,r)$ the expansion of $l^{a}$ is zero. We can also compute the value $n^{a}\nabla_{a}\theta_{l}$ at $r=2m$
\beq \left(n^{a}\nabla_{a}\theta_{l}\right)_{H} = - \frac{(1-2m'_{H})}{r_{H}^{2}}\left(1+\frac{\dot{r}_{H}}{2e^{-\Phi_{H}}}\right) \eeq
where will we use a dash to denote partial derivative with respect to $r$ and a dot to denote the partial derivative with respect to the time $\tau$ (here, since $r_{H}$ is only a function of $\tau$ it is actually an ordinary derivative).

For the horizon to be an outer horizon we require $2m'_{H} < 1$, since $m(\tau,r)$ must be less than $r$ for large $r$. In addition, we can see from (\ref{dtdtau}) for the ingoing null geodesic $n^{a}$ that $\dot{r} = -2e^{-\Phi_{H}}$. Thus we see that we have a trapping horizon at $r=2m$ if the horizon is outer and not moving inwards faster than ingoing null geodesics.

The normal $N^{a}$ to the surface $r=2m$ has norm
\beq N^{a}N_{a} = -4\dot{m}e^{2\Phi}-4\dot{m}e^{\Phi}(1-2m') \eeq
If $\dot{m}=0$ the trapping horizon will be a null hypersurface, and, assuming $1-2m'>0$, it will be a spacelike hypersurface if $\dot{m}>0$. For $-(1-2m')e^{\Phi} < \dot{m}<0$ the trapping horizon will be a timelike hypersurface. This opens the possibility that one can move along a causal curve from inside an evaporating horizon to the outside. For $\dot{m}<-(1-2m')e^{\Phi}$ the horizon is spacelike, but evaporating `faster than the speed of light' and so all timelike curves from a region just inside the horizon must move to the outside \cite{Nielsen:2005af}.

The surface $r=2m(r,t)$ does not however, define the location of the event horizon in a dynamical spacetime. The event horizon is always a null surface and so the spherically symmetric trapping horizon at $r=2m$ can only be an event horizon if $\dot{m}=0$ (note however that this is necessary but not sufficient). To find the event horizon, firstly one would need an explicit form for $m(r,t)$ and then one would look for radial null vectors that are not able to reach infinity by propagating them outwards from the centre of the spacetime. 

\section{Thermodynamics for Trapping Horizons}

Dynamical laws analogous to the usual laws of thermodynamics can easily be derived for the above spherically symmetric trapping horizons. As shown above, the surface defined by
\beq \label{surfcond} r=2m(\tau,r), \eeq
defines a trapping horizon in many cases. Differentiating this equation with respect to any parameter $\xi$ labeling spherically symmetric foliations of the horizon, gives
\beq \frac{\d r}{\d\xi} = 2\frac{\partial m}{\partial \tau}\frac{\d \tau}{\d\xi} + 2\frac{\partial m}{\partial r}\frac{\d r}{\d\xi}. \eeq
If we take $\xi = \tau$ and rearrange using the formula for the area $A=4\pi r^2$ this becomes
\beq  \label{firstlaw} \frac{\partial m}{\partial \tau} = \frac{1}{8\pi}\frac{(1-2m')}{2r}\frac{\d A}{\d \tau}, \eeq
where $m'=\frac{\partial m}{\partial r}$. In order for this to take the same form as the first law of black hole thermodynamics $\d m = \frac{1}{8\pi}\kappa\;\d A$ it seems natural to take
\beq \label{Nsurfgrav} \kappa_{N} = \frac{(1-2m')}{2r_{H}}. \eeq

as a definition of surface gravity, defined by the first law and normalised by the choice of quasi-local mass, in this case the Misner-Sharp mass \cite{Nielsen:2007ac}.\bigskip

In order to obtain a version of the second law we can just compute $G_{ab}l^{a}l^{b}$, where $G_{ab}$ is the Einstein tensor. This gives
\beq G_{ab}l^{a}l^{b} = \frac{2e^{\Phi}}{r^{2}}\frac{\partial m}{\partial \tau}\sqrt{\frac{2m}{r}} - \frac{2}{r}\frac{\partial\Phi}{\partial r}\left(1-\sqrt{\frac{2m}{r}}\right)^{2}. \eeq
Rearranging gives
\bea \frac{\partial m}{\partial \tau} & = & \frac{1}{2}e^{-\Phi}r^{2}\sqrt{\frac{2m}{r}}G_{ab}l^{a}l^{b} + \nonumber \\ & & e^{-\Phi}\Phi'\sqrt{2mr}\left(1-\sqrt{\frac{2m}{r}}\right)^{2} \eea
At $r=2m$ we can impose (\ref{firstlaw}) and so we find
\beq \frac{\partial A}{\partial \tau} = \frac{16\pi r^{3}e^{-\Phi}}{1-2m'} G_{ab}l^{a}l^{b} \eeq
Thus we see that the area of the horizon $A$ is increasing if $G_{ab}l^{a}l^{b} > 0$. By the Einstein equations we can write this condition as $T_{ab}l^{a}l^{b} > 0$, which is exactly as we expect. The area of the horizon is increasing if the null energy condition is satisfied, the area of the horizon is constant if the null energy condition is saturated and can decrease only if the null energy condition is violated.

This is the only place where the Einstein equations come into play. Since the derivations are only based on the behaviour of what is essentially a `metric ansatz', all the other results should apply to an arbitrary matter theory with arbitrary curvature corrections.

The above laws of black hole mechanics are of course coordinate dependant in that they depend on the time parameter $\tau$ (and the radial coordinate $r$). However, there is nothing particularly special about this choice of time parameter. Any good parameter on the horizon will give similar laws of mechanics. What is essential is that these laws hold at $r=2m$, which as we have seen above, defines a trapping horizon and in general, does not define the event horizon.

\section{Hawking radiation for Trapping Horizons}

If trapping horizons can give rise the thermodynamic laws just like event horizons, which horizon should be associated with `true' thermodynamic behaviour? Di Criscienzo et al. have investigated the production of Hawking radiation by trapping horizons \cite{Di Criscienzo:2007fm}. Similar results have been obtained earlier by Visser in \cite{Visser:2001kq}. We will give a brief recap of the argument. Consider the equation for a massless scalar field on a curved background
\beq \frac{\hbar^{2}}{\sqrt{-g}}\partial_{a}\left( g^{ab}\sqrt{-g}\partial_{b}\right)\phi = 0 \eeq
We look for solutions of the form $\phi = exp(-iS(\tau, r)/\hbar)$ (we ignore the amplitude). Taking the limit as $\hbar \rightarrow 0$, to lowest order this equation gives the Hamilton-Jacobi equation
\beq \label{hamiltonjacobi} g^{ab}\partial_{a}S\partial_{b}S = 0 \eeq
Invoking the geometrical optics approximation, which will be valid when the wavelength is small with respect to the curvature and is changing slowly on a scale with respect to the frequency,
\beq S(\tau,r) = \omega t - \int k(r)\d r \eeq
equation (\ref{hamiltonjacobi}) gives
\beq \omega^{2} + 2e^{-\Phi}\sqrt{\frac{2m}{r}}\omega k - e^{-2\Phi}\left(1-\frac{2m}{r}\right)k^{2} = 0 \eeq
Solving quadratically for $k$ gives
\beq k = \pm\frac{\omega e^{\Phi}}{1\mp\sqrt{\frac{2m}{r}}} \eeq
The upper sign denotes the outgoing modes and the lower sign denotes the ingoing modes. The outgoing modes contain a simple pole at $r=2m$, the location of the trapping horizon. We can examine the contribution to the phase $S$ of the outgoing modes by expanding around the horizon.
\beq S = \omega t + \frac{2r_{H}\omega e^{\Phi_{H}}}{\left(1-2m'_{H}\right)}\int\frac{\d r}{\left(r-r_{H}\right)} \eeq
This integral can be performed by deforming the contour into the lower half of the complex plane, which gives a complex contribution to $S$
\beq \textrm{Im}S = \frac{4\pi r_{H}\omega e^{\Phi_{H}}}{\left(1-2m'_{H}\right)} \eeq
It is well known that this calculation gives rise to a tunneling probability of
\beq \Gamma \sim \phi\phi^{*} = e^{-2\mathrm{Im}\; S} \eeq
For a thermal spectrum we expect a tunneling rate proportional to a Boltzmann factor $\Gamma \sim e^{-\omega/T}$. At this level of approximation this corresponds to thermal radiation with a temperature
\beq T = \frac{1}{2\pi}\frac{e^{-\Phi_{H}}}{2r_{H}}\left(1-2m'_{H}\right) \eeq
which agrees with the calculations in \cite{Nielsen:2007ac}.

This seems to suggest that it is exactly the pole at $r=2m$ that is responsible for the tunnelling flux through the horizon. This is of course the trapping horizon and not the event horizon. Now this is far from being incontrovertible proof that a trapping horizon is required for Hawking radiation. But it is a least suggestive that it may have some role to play and further research may clarify the picture.

\section{Gravitational entropy}

If the Hawking radiation is to be associated with the trapping horizon, one can ask what about the gravitational entropy. There are a number of reasons why one might associate entropy to black holes. The first, considered by Wheeler, was the apparent unverifiability of the second law of thermodynamics if objects such as hot and cold tea were dropped into a black hole \cite{Wheeler:1998vs}. This led Bekenstein to postulate that the area of a black hole should be seen as a measure of the interior state of the black hole that is inaccessible to an external observer \cite{Bekenstein:1973ur}.

Furthermore, Hawking showed that a black hole could lead to a breakdown of predictability since taking the trace over the unknowable interior state would turn an initial pure quantum state into a thermal state \cite{Hawking:1976ra}. All three of these arguments would seem to suggest that an entropy can only truly be associated to event horizons, and not trapping horizons. As we have seen above the trapping horizon is not always a null surface and when its area is decreasing it will be a timelike surface, allowing causal signals to propagate across it in both directions. In spacetimes without event horizons the state of the interior of a trapping horizon black hole may eventually become accessible to outside observers \cite{Hayward:2005gi,Ashtekar:2005cj}.

Another reason to associate entropy with black holes is that several models for quantum gravity have been able to count the microstates that give rise to this entropy. It is interesting in this context to note that this has only been shown in Loop Quantum Gravity for isolated horizons, which while locally defined in a fashion similar to trapping horizons, have no true dynamics and thus appear very similar to stationary event horizons. The fuzzball picture in string theory seems to describe an object with no true event horizon \cite{Mathur:2005zp}. 

\section{Conclusion}

We have seen above that there remains much to be discovered about black holes in four dimensions, even in semi-classical theories with ordinary matter fields. Trapping horizons may offer some insights into puzzling black hole behaviour. However, much remains to be done to establish them as truly viable definitions for black holes. A common criticism of apparent horizons and marginally trapped surfaces (on which the definitions of trapping horizons are based) is that their existence and position depends on a choice of foliation. In the Schwarzschild solution for example, different foliations have apparent horizons in different places and some foliations have none at all \cite{WaldandIyer}. To counter this, one can take the attitude that if there exists a hypersurface which admits the structure of a future outer trapping horizon then the spacetime can be said to contain a black hole. The uniqueness of the future outer trapping horizon structure also remains and open question. One may have to take the attitude that one defines the black hole region only by the outermost horizon.

Another question is whether we really can do without event horizons. As we have seen, many of the arguments in favour of assigning gravitational entropy to black holes seem to apply best to event horizons. In general relativity we have the celebrated singularity theorems that imply that trapped surfaces (which are closely related to trapping horizons) lead to singularities. By the cosmic censorship hypothesis, one can then argue that such singularities should be covered by event horizons. This seems to imply that trapping horizons will always be associated with event horizons. However, these theorems depend on the hypothesis that our universe can always be described by a smooth manifold and secondly the singularity theorems depend on an energy condition such as the null energy condition, which we expect to be violated if Hawking radiation can occur. We have also seen above that the question of whether event horizons truly exist in our universe or not is almost impossible to determine experimentally. It may well be that if one wants to restrict oneself to studying the physics of our universe one must do without event horizons.

We end with some open speculation. A universe that contains trapping horizons but no true event horizons appears to be a possibility. Such a spacetime would be able to account for all astrophysical observations of black holes and may well exhibit Hawking radiation. Whether our universe is such a universe may not be answerable. But it seems that the close analogy between the laws of thermodynamics and black holes, that was clinched by the discovery of Hawking radiation, may not be so exact after all. If the Hawking radiation does not derive from the event horizon, but the gravitational entropy does, then these two phenomena must be seen as logically separate. In addition, the temperature of the black hole that one can compute from the form of the first law of black holes mechanics does not match entirely the temperature that we have calculated for the Hawking flux using the tunneling approach \cite{Nielsen:2007ac}.

It is not so much the matter that falls into the black hole that evaporates, but rather the spacetime itself. The thermality of the Hawking radiation is not related to lost information about the matter that originally formed the black hole, but rather solely to the state of the gravitational field that, in a certain sense, at least in vacuum, generates itself. This is similar to the view that the gravitational field of the Schwarzschild solution is caused, not by some infinitely dense source of mass at $r=0$, but by the non-linear self-coupling of the gravitational field to itself. Therefore one should not expect correlations between the infalling matter and the outgoing Hawking radiation, unless these correlations can be measured in the gravitational field itself.

Of course none of this answers the question of what happens to the infalling matter as it reaches the central singular region. For that one really would need a microscopic description of spacetime and particles. One would need to know what replaces the singularity in a full theory of quantum gravity. And this would also be needed to predict what happens when the evaporating black hole becomes very small and approaches this `singular' region. A true understanding of what black holes and black hole evaporation can teach us about the form such a theory of quantum gravity will take, may require some reevaluation of our current notions.

\end{document}